\documentclass[aps,pre,reprint,onecolumn,superscriptaddress]{revtex4-2}
\def\rtx@apsprf{
\class@info{APS journal PRF selected}
}
\usepackage{graphicx}
\usepackage{epstopdf,epsfig}
\usepackage[utf8x]{inputenc}
\usepackage{afterpage}
\usepackage{graphicx}
\usepackage{hyperref}
\usepackage[english]{babel}
\usepackage{color}
\usepackage{dcolumn}
\usepackage{bm}
\usepackage{amsfonts}
\usepackage{amsmath}
\usepackage{amssymb}
\usepackage{amsbsy}
\usepackage{subfigure}
\usepackage{psfrag}
\usepackage{float}
\usepackage{lineno}
\usepackage{layouts}
\usepackage{epstopdf}
\usepackage{lineno}
\usepackage[normalem]{ulem}
\usepackage{tabularx,booktabs}
\usepackage{array}
\usepackage{comment}

\usepackage{graphicx}
\usepackage{dcolumn}
\usepackage{bm}
\usepackage{color-edits}
\usepackage[dvipsnames]{xcolor}
\addauthor[Juan]{jd}{MidnightBlue}
\addauthor[Manuel]{mld}{ForestGreen}
\begin{document}

\title{D-shaped body wake control through flexible filaments}

\author{J. C. Mu\~noz-Herv\'as}
\affiliation{Departamento de Ingenier\'{\i}a Mec\'anica y Minera, Universidad de Ja\'en, Jaén, Spain}
\affiliation{Andalusian Institute for Earth System Research, Universities of Granada, Jaén y Córdoba, Spain.}

\author{B. Semin}
\affiliation{Laboratoire PMMH, ESPCI, CNRS, Sorbonne Universit\'e, Universit\'e Paris Cit\'e. Paris, France.}
\author{M. Lorite-D\'iez}
\email{mldiez@ugr.es}
\affiliation{Andalusian Institute for Earth System Research, Universities of Granada, Jaén y Córdoba, Spain.}
\affiliation{Departamento de Mec\'anica de Estructuras e Ingenier\'ia Hidr\'aulica, Universidad de Granada, Spain.}
\author{G. J. Michon}
\affiliation{Institut Jean Le Rond d'Alembert, CNRS, Sorbonne Universit\'e, UMR 7190, F-75005 Paris, France}
\author{Juan D'Adamo}
\affiliation{Facultad de Ingenier\'ia Universidad de Buenos Aires, CONICET. Buenos Aires, Argentina.}
\affiliation{Institut Franco-Argentin de Dynamique des Fluides pour l'Environnement, IRL2027, CNRS, Universidad de Buenos Aires, CONICET. Buenos Aires, Argentina.}
\author{J.I. Jim\'enez-Gonz\'alez}
\affiliation{Departamento de Ingenier\'{\i}a Mec\'anica y Minera, Universidad de Ja\'en, Jaén, Spain}
\affiliation{Andalusian Institute for Earth System Research, Universities of Granada, Jaén y Córdoba, Spain.}
\author{R. Godoy-Diana}
\affiliation{Laboratoire PMMH, ESPCI, CNRS, Sorbonne Universit\'e, Universit\'e Paris Cit\'e. Paris, France.}

\date{\today}

\begin{abstract}
In this study, we investigate the flow around a canonical blunt body, specifically a D-shaped body of width $D$, in a closed water channel. Our goal is to explore near-wake flow modifications when a series of rigid and flexible plates ($l=1.8D$) divided into filaments ($h=0.2D$) are added. We focus on assessing the interaction between the flexible filaments and the wake dynamics, with the aim of reducing the recirculation bubble and decreasing the velocity deficit in the wake. To achieve this, we conduct a comparative study varying the stiffness and position of the filaments at different flow velocities. 
The study combines Particle Image Velocimetry (PIV) measurements in the wake behind the body with recordings of the deformation of the flexible filaments. Our observations show that the flexible filaments can passively reconfigure in a two-dimensional fashion, with a mean tip deflection angle that increases with the incoming flow velocity. Deflection angles up to approximately $\sim 9^\circ$ and vibration tip amplitude of around $\sim 4^\circ$ are achieved for flow velocities  $U^{*}\simeq f_{n}D/u_{\infty}\geq 1.77$, where $f_n$ is the natural frequency of the flexible filaments. This reconfiguration results in a reduction of the recirculation bubble and a decrease in the velocity deficit in the wake compared to the reference and rigid cases. In addition, curved filaments with a prescribed rigid deformation exhibit very similar behavior to that of flexible filaments, indicating that the vibration of flexible filaments does not significantly disturb the wake. The obtained results highlight the interest of testing flexible appendages in the wake of blunt bodies for designing effective flow control devices.

\end{abstract}
\maketitle
\section{Introduction}\label{sec:Intro}
The study of the flow around two-dimensional blunt bodies is a significant research field, as this configuration is common in many engineering and environmental applications, including civil structures such as bridges and buildings, as well as offshore structures like deep-water risers. In particular, flow separation at the rear of these bodies can generate vortex shedding, caused by Bénard-von Karman instability (BvK), leading to the formation of a low-pressure region. Research in this area is typically driven by practical goals such as reducing drag on the blunt body \cite{Choi2008}, suppressing flow-induced vibrations (FIVs) \cite{Williamson2004, Sarpkaya2004, Paidoussis2010}, or modifying heat transfer around a body immersed in a given flow \cite{Lecordier1991}.


In the study of two-dimensional bluff bodies, D-shaped geometries have been extensively analyzed, particularly due to their characteristic fixed separation point at the trailing edge \cite{Tombazis1997, Park2006, Pastoor2008}. For these geometries, the mean base pressure coefficient remains nearly constant along the third direction, with values around -0.6. This constancy emphasizes the wake’s significant impact on overall drag, which ranges between 0.8 and 1 for the tested Reynolds numbers. The Reynolds number, defined as $Re = u_{\infty}D/\nu$, depends on the free-stream velocity ($u_\infty$) the body diameter ($D$) and the fluid's kinematic viscosity ($\nu$). 
In this configuration, vortex shedding occurs at a specific frequency, $f$, or equivalently, at a fixed Strouhal number $St=fD/u_{\infty}$, as for instance, $St=0.25$ at $Re=40.000$.  The shedding process produces strong velocity fluctuations in the wake downstream of the body \cite{Park2006} that impact the body aerodynamics.

Various devices and strategies aim to actively or passively control the wake behind blunt bodies to reduce the averaged or fluctuating drag or lift force coefficients. Active control devices include pulsed jets \cite{Parkin2014, Li2019}, plasma actuators \cite{Boury2018}, or suction/blowing strategies \cite{Bearman67, Sevilla04}. Examples of closed-loop flow control include sinusoidal zero-net-flux actuation through spanwise slots \cite{Henning2007, Pastoor2008} and actuated flaps in 3D wakes \cite{Brackston16}. However, implementing active flow control systems in real-world applications often requires significant power input or sophisticated systems, limiting their practicality and efficiency. This makes passive control systems more attractive due to their simplicity.


Among passive strategies, initial studies by \cite{Mair65} examined the effect of including plates at the body base, while other more recent studies introduced trailing edge modifications such as small tabs \cite{Park2006}, base cavities \cite{Sanmiguel2011, Cai2009, Kruiswyk1990} and boat-tailed after-bodies \cite{Mair1978, Han1992, Choi14}. Rear cavity devices can be further refined with multi-cavity arrangements \cite{Martin2014} or optimized geometries focused on reducing drag \cite{Lorite17}. The efficiency of rear cavities in terms of drag reduction is linked to their depth, as they move the wake formation away from the body base, generating a dead flow region that effectively increases base pressure.

However, their aerodynamic efficiency may be reduced under different flow conditions, such as accelerated flows from rest \cite{Lorite2018} or yawed orientations \cite{Lorite20a}. An interesting approach that can adapt to different flow conditions involves using bio-inspired flexible devices, modifying the rear part of the body into a more aerodynamic shape through passive reconfiguration \cite{Alben2002, Garcia2021, Garcia2023}. This process is commonly observed in nature, where the reconfiguration of flexible parts \cite{Harder2004, Zhang2020}, such as leaves in trees, reduces aerodynamic resistance and effective area \cite{Vogel1989} under different wind conditions, (see \cite{Langre2008}, for a review). The passive reconfiguration of flexible systems to modify the wake behind bluff bodies has been only scarcely investigated \cite{Abdi2019}. Recent studies \cite{ Garcia2021, Garcia2023} propose using rotary and flexible flaps to configure a self-adaptive cavity, achieving drag reductions of up to 18$\%$ by reducing the wake bluffness and streamlining rear separation.
 
Despite these advantages, the practical implementation of these devices may face challenges, such as ensuring uniform mechanical properties in movable parts with large length-to-thickness ratios and managing significant loads and potential torsional deformations. As an alternative, we propose using a modular system based on an array of small filaments that can function as a single flexible flap with much lower mechanical loads. In the case of rigid arrays of filaments, \cite{Keirsbulck2023} examined the use of porous rear cavities in 3D blunt bodies, finding similar results for modular cavities with small spacing between elements and fully solid rear cavities. Additionally, the device studied here incorporates the degrees of freedom associated with the flexibility of the filaments, allowing the elements to adapt their shape to the eventual three-dimensional features of the flow around the D-shaped body.

In this work, we aim to study the effect of modular arrays of filaments resembling a cavity to manipulate and control the wake behind a D-shaped body. The tested devices comprise an array of flexible filaments that interact with the incoming flow, adapting their shape to different conditions. We will experimentally assess the interaction between the filaments and the wake by measuring filament motion and wake features. The experimental details are presented in Sect.  \ref{sec:Experiment}, followed by the main results in Sect.\ref{sec:Results}. There, Sect. \ref{subsec:Response} describes the flexible filaments motion, Sect. \ref{subsec:FSI} covers the temporal synchronization between filaments and vortex shedding, and Sect. \ref{subsec:Wakechange} presents the wake features for the four tested configurations. The main conclusions are drawn in Sect. \ref{sec:Conclusions}.

\section{Experimental setup}\label{sec:Experiment}
\begin{figure}[t]
\centering
\includegraphics[width=1\textwidth]{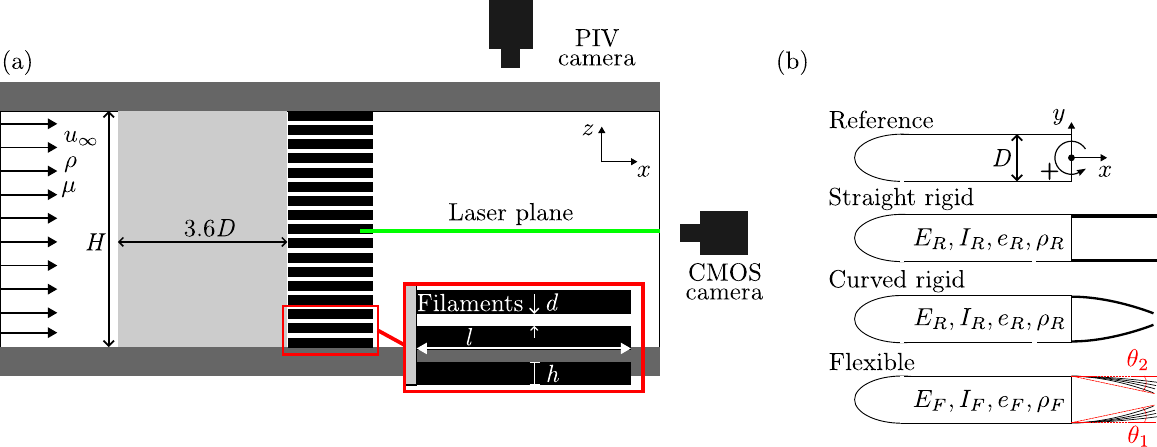}
\caption{(a) Sketch of the experimental set-up and (b) Tested rear configurations and their corresponding characteristic parameters.}
\label{Fig1Sketch}
\end{figure}

The experiments were carried out in a closed water channel with a cross section of 0.1 $\times$ 0.1 $\mbox{m}^2$ and a turbulence intensity below  3$\%$ (see Fig.\ref{Fig1Sketch} a). This water channel has been characterized and used in previous studies focused on 3D bluff body wakes as \cite{Chrust2013, Klotz2014, Skarysz2018}. Here, the D-shaped body used has a characteristic diameter $D$ of 20 mm, a height $H = 5D$ and length $L = 3.6D$. The blockage effect and the development of boundary layers in the channel walls have been taken into account in our study, appropriately re-scaling the results for all flow conditions tested as in \citep{Marais2011}. In our experiment, the coordinate system is centered at the body base, with the $x$-axis aligned with the streamwise direction, the $z$-axis pointing vertically, and the $y$-axis completing a right-hand trihedron.

In that reference frame, the rotations in the anti-clockwise direction are considered positive. Four different configurations are tested and compared, including the reference body with blunt trailing edges (reference), and three configurations that contain two arrays of filaments with different stiffness and shapes implemented in the trailing edges of the D-shaped body. Each array contains 17 filaments, with length $l=1.8D$ and height $h=0.2D$. Each filament is separated from the following one by $d=0.1D$ so the filaments are free to move separately. 
The filaments are made of TPU (Thermoplastic Polyurethane) 92A with an elastic modulus of $E_{F} =16$ MPa (Flexible) and made of PLA (polylactic acid) with $E_{R}=3600$ MPa (Rigid). In the case of rigid arrays of filaments, we have tested two shapes of filaments, one aligned with the body trailing edges (straight) and one with a given curved deformation (curved), corresponding to a tip deflection angle of $11^{\circ}$ for a first Euler-Bernoulli vibration mode. This reconfiguration angle is the maximum reconfiguration observed for a single flexible filament in our experiment. The height of the body has been adjusted so that the filaments can move freely without touching the walls of the channel. However, as will be explained in subsequent sections, it has been observed that the upper and lower filaments are affected by the proximity of the walls.


The arrays of filaments were first designed through CAD and then manufactured using 3D printers. In particular, the flexible filaments were produced with an Artillery X2 printer using TPU (Thermoplastic polyurethane), specifically TPU-92A from Kimya. The number 92A refers to the hardness in shore scale, the reported young modulus (16 MPa) from the vendor does not differ from our measurements. For the rigid filaments, a printer Stratasys Fortus 250mc was employed using PLA. In addition to changing the employed material, different filament thicknesses were used to manufacture appropriate rigid (not interacting with flow) and flexible filaments. For the flexible ones, a thickness, $e_{F} = 0.01D$ is used, while for the straight and curved ones,  thickness, $e_{R} = 0.05D$ is employed. Regarding the flexible filaments, the dimensions were meticulously selected to achieve sufficient reconfiguration capacity, with numerous tests conducted using different lengths and thicknesses.


The employed water channel works in a range of free-stream velocities ranging between 0.05 and 0.28 m/s, corresponding with Reynolds numbers from 1060 to 5200. Accordingly, the reduced velocity, defined as the ratio between the free-stream velocity and the averaged velocity of the flexible filaments movement, $U^* = u_{\infty}/f_{n}D$ results within the interval $ [0.5, 2.45]$,  The latter is  determined by the flexible filaments corresponding natural frequency. This natural frequency in vacuum has been firstly obtained with the following equation:

\begin{equation}
    f_n=\frac{\beta^2}{2\pi}\sqrt{\frac{E I}{\rho_F e_F h l^4}},
\end{equation}
where $\beta$ is a coefficient given by the Euler-Bernoulli deflection mode for a cantilevered beam, $I$ the moment of inertia of one filament around the bending axis and $\rho_F$ the density of the TPU 92A, which is the material employed for the flexible filament array. The obtained result ($f_n=5.7~Hz$) matched accurately the natural frequency measured in free-decay tests ($f_n=5.67~Hz$) made in air.  These tests were conducted using a position laser sensor and initially displacing the filaments from their equilibrium position. Note that no valid free-decay tests are available for plates in water, as the motion is quickly damped; nevertheless, a simplified theoretical estimation of the natural frequency in water, following \cite{Garcia2023}, provides a value of $f_{n,w} \leq f_{n}/3$. This approximation does not consider any added mass coefficient and therefore, its use for a deeper analysis of results has been discarded. 

We carried out PIV measurements on the 4 employed configurations to analyze the wake behind the D-shaped body in the $xy$ plane for each case. To generate the illuminated measurement plane, we horizontally positioned at $z=0$ a laser sheet generated by a Quantel Big Sky double pulsed Nd:YAG laser with 532nm emision with 250mJ per pulse at 10 Hz, as shown in Fig.\ref{Fig1Sketch} (a). That laser illuminated polyamide particles with a characteristic diameter of 20 $\mu$m (Lavision HQ polyamide beads). To capture the seeding particle images, we employed a PIV system comprising a LaVision{\textregistered} PTU-X synchronizer and a SCMOS Imager Pro Plus 2M 12-bits camera  equipped with a Nikon 50mm objective and an aperture fixed at f/5.6 (see PIV Camera in Fig.\ref{Fig1Sketch}). The resulting images were 1600 $\times$ 1200 pixels and were calibrated by measuring the distance between the channel walls, yielding a magnification factor of 10.58 px/mm. The recordings were conducted at a frequency of 10 Hz with laser pulses delayed by approximately 0.7 to 3  ms, capturing a total duration of 60 s (600 pairs of images per experiment). After correctly setting the region of interest (123 $\times$ 87 mm), velocity vectors were obtained from cross-correlation in Davis 10 from LaVision{\textregistered} with interrogation windows of 32 $\times$ 32 pixels with an overlap of 50$\%$, resulting in a spatial grid of $83 \times 59$  points, with a resolution of 13 vectors in the body diameter, $D$. Furthermore, to ensure the repeatability and the accuracy of the results, different PIV tests were conducted at the same conditions in different days, without observing any significant difference between tests. Finally, our measurements have a correlation-based uncertainty \citep{Wieneke2015} typically below $3\%$ of the free-stream velocity for both $u_x$ and $u_y$.

In addition, for the flexible filaments, after characterizing their mechanical properties, we recorded their tip deflection with a Basler ACE acA2000-165um 	CMOS camera, equipped with a CMV2000 Sensor, which is capable of acquiring 2048 $\times$ 2048 pixel images at a maximum frame rate of 165 fps. Thanks to the available optical access at one end of the water channel, this camera was able to characterize the filaments motion in the $zy$ plane. In order to track this motion, each filament edge was highlighted with white dots which were illuminated by a high-power LED lamp. The camera recordings were made at 134 fps and the exposure time was adjusted to capture static white dots.

To analyze the captured images, we employed the open-source image processing tool DLTdv8 \citep{Hedrick2008}, which can be integrated as a toolbox within Matlab{\textregistered}. This tool allows precise tracking of specific points in a video. Initially, it conducts an image analysis based on a predetermined threshold, followed by contour detection facilitated by intensity discrepancies.

In terms of the experimental measurements reliability and precision, the deformation measurements were repeated at two specific $Re$ numbers (1196 and 3635), observing errors below 5$\%$ of the reported values, the corresponding error bars are also included in the figures. 

 
In the following, time-dependent variables will be denoted using lower-case letters $a$, while time-averaged values will be expressed with upper-case letters $A=\overline{a}$. In addition, $\hat{a}$ will denote the instantaneous fluctuating amplitude of the variable $a$, which will be computed by means of the Hilbert transform, so that $\hat{A}$ will represent the corresponding time-averaged fluctuating amplitude. On the other hand, the notation $\mid A \mid$ will be used to indicate the modulus of a variable. Finally, $\langle a \rangle$ stands for the spatial average of $a$ variable. 

\section{Results}\label{sec:Results}

\subsection{Response of the flexible filament array}\label{subsec:Response}

\begin{figure}[h!]
\centering
\includegraphics[width=0.6\textwidth]{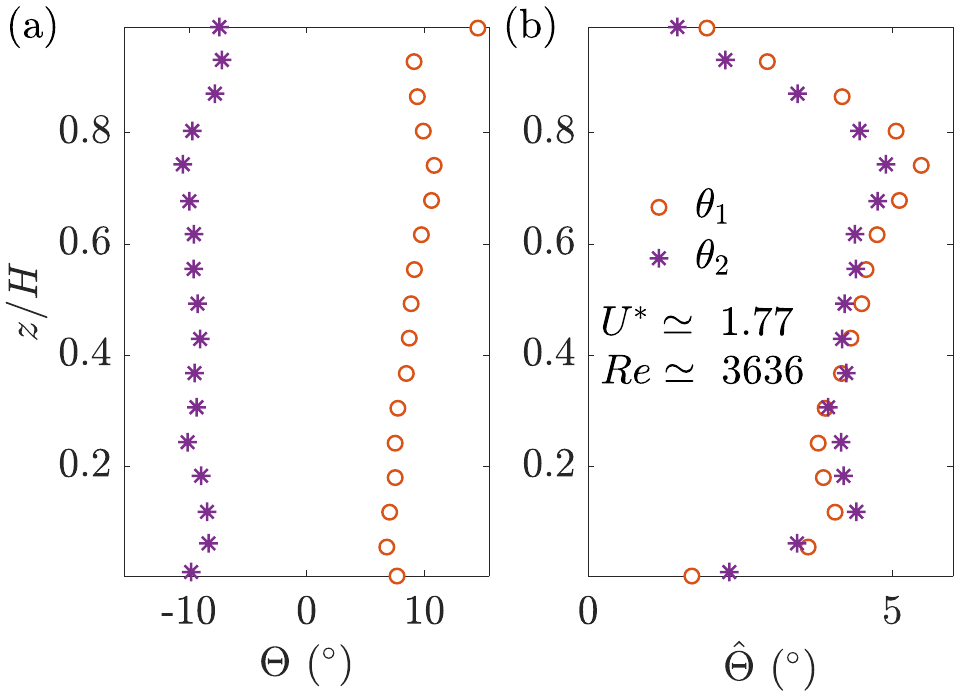}
\caption{(a, b) Time-averaged tip angle deflection, $\Theta$, and the corresponding averaged oscillations amplitude, $\hat\Theta$, for the flexible filaments along the spanwise direction, of the D-shaped body at $Re\simeq3636$ and $U^{*}\simeq 1.77$. We depict in orange circles the filaments on the left side ($y<0,~\theta_1$) and in purple asterisks those on the right side ($y>0,~\theta_2$).}
\label{Fig2AngZ}
\end{figure}

We will firstly analyze the motion of the two arrays of filaments.  We employed a rear view to capture the tip deflection of total 34 flexible filaments (17 on each  side). The tip angles, $\theta$, were obtained with the filament length, $l$, by referring them to the reference body trailing edge position. We also checked that the filaments were deformed according to a first Euler-Bernoulli deflection mode for a cantilevered beam by supplementary recordings from the top of the water channel (the recording location is similar to the one of the PIV camera). The analysis of the captured images depicts that all filaments bend on average inward the wake while describing a periodic fashion once a certain flow velocity is set. They move in phase along the span as well as between the two arrays of filaments. The analysis of the temporal deformation of the filaments provides both the mean deflection angle $\Theta$ and the angular oscillation amplitude $\hat{\Theta}$ for each filament, whose spanwise distributions for $U^{*}\simeq1.77$ $(Re\simeq 3636)$ are shown in Fig. \ref{Fig2AngZ}. The filaments on the left ($y<0$), $\theta_1$, and right ($y>0$), $\theta_2$, sides are shown along the spanwise direction, $z$. Figure \ref{Fig2AngZ} (a) shows a clear reaction in a quasi-static 2D way of the flexible filament arrays under the influence of the flow. At the depicted $U^{*}$, the flexible filaments reach an averaged tip deflection around $9^{\circ}$ along with moderate oscillations of approximately $4^{\circ}$ (see Fig. \ref{Fig2AngZ}b). The spanwise distribution of the flexible filaments deflections and oscillations shows a seemingly constant behaviour in the the central part of the D-shaped body with distinguishable differences close to the water channel walls ($z/H < 0.15$, $z/H>0.85$). Apart from the wall influence, the response of the arrays of filaments could be, in a first approximation, analyzed as a 2D response. 

\begin{figure}[h!]
\centering
\includegraphics[width=0.9\textwidth]{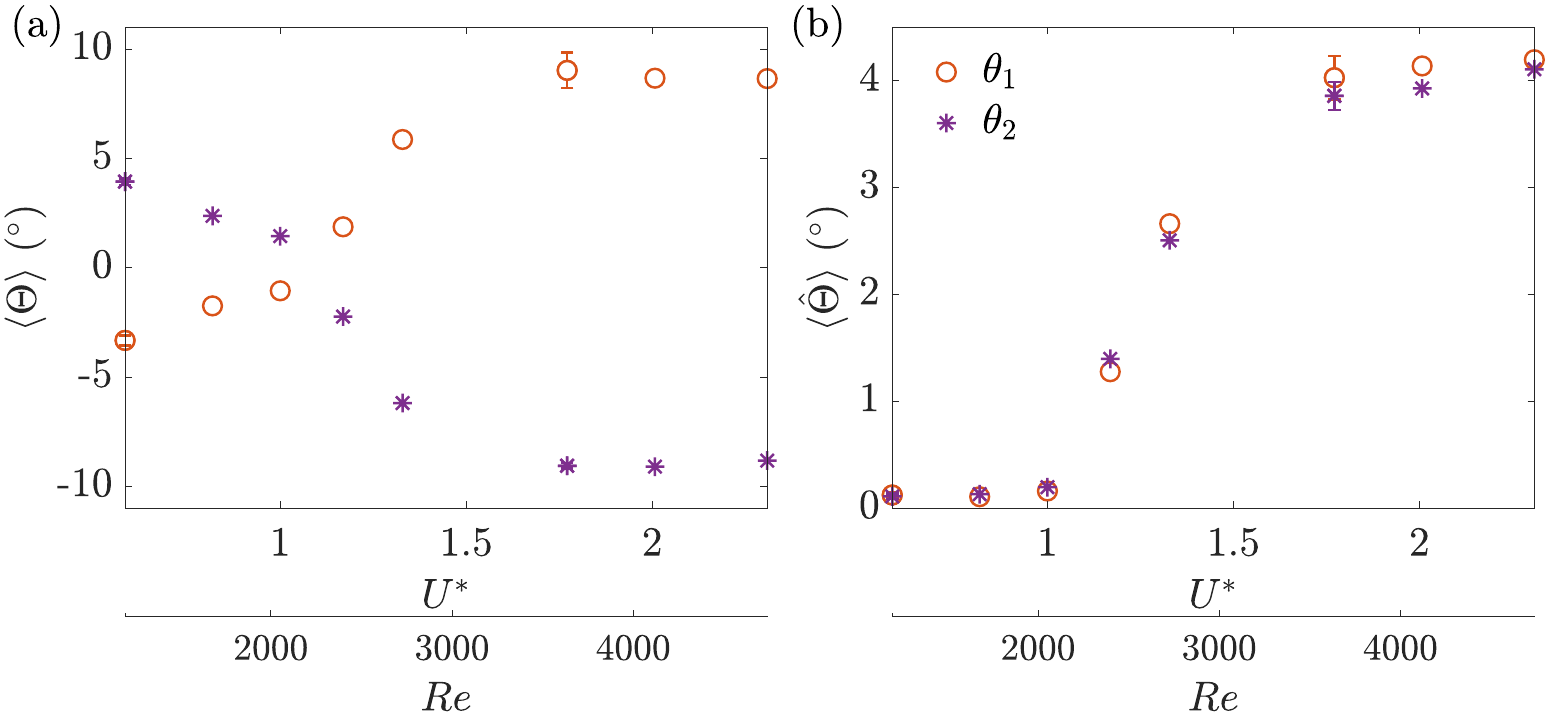}
\caption{Evolution of (a) the spatio-temporal averaged tip deflection angle, $\langle\Theta\rangle$, and (b) the  spatio-temporal averaged amplitude of tip angle oscillations, $\langle\hat \Theta\rangle$, with respect to $U^*$ and $Re$. We depict in orange circles the filaments on the left side $(y<0,~\theta_1)$, and in purple asterisks those on the right side $(y>0,~\theta_2)$. 
}
\label{Fig3MAng}
\end{figure}

The evolution of the mean tip deflection angle, averaged in time and in space across the spanwise direction, is presented in Fig. \ref{Fig3MAng} (a) for the range of $U^{*}$ ($Re$) tested here.  The mean deflection, $\langle\Theta\rangle$, increases with increasing Reynolds numbers and reduced velocities until reaching a saturation value, around $9^\circ$, at $U^*\simeq1.77$ or $Re\simeq3636$. This absolute increase of the deflection shown by $\langle\Theta\rangle$, has an opposite sign for each array of filaments, due to the sign convention employed, where a rotating angle in the counterclockwise direction is positive (see Fig. \ref{Fig1Sketch}b). Thus, the filaments on the left side ($\theta_1$) have increasingly positive angles as filaments deflection increases with $U^{*}$ whereas the ones on the right side ($\theta_2$), depict increasingly negative angles with increasing $U^{*}$. Furthermore, it is noteworthy that there is an initial part of the range, until $U^{*}<1$ and $Re<2052$, an initial deformation in the outward direction is observed for both flaps, meaning that the position of the filament tips extends beyond the limits of the D-shaped body. This is why both data series cross $\langle\Theta\rangle = 0$ immediately after $U^{*}\simeq1$ ($Re\simeq2052$). In absolute terms, this region with a negative mean deflection indicates that the flow momentum is insufficient to deform and move the filaments and the filaments preserve their initial random dry loading. The evolution of the mean tip deflection angle depicted in Fig. \ref{Fig3MAng} (a) illustrates an increasing passive reconfiguration of the flexible filaments in the studied range of $U^*$ ($Re$) for both arrays of flexible filaments for reduced velocities above $U^{*}>1$, what may lead to efficient reduction of the drag coefficient. However, the passive reconfiguration reaches a saturated state around $\Theta=9^\circ$ for $U^{*}>1.77$, suggesting a limit on the potential for drag reduction from this point onward, that may translate into an asymptotic value of the Vogel exponent \citep[see e.g.][]{Garcia21} (note that such a negative exponent is typically used to quantify the aerodynamic effect of passive reconfiguration of flexible structures \cite{Vogel1989, Garcia2023}).

Regarding the filaments vibrations, the spatio-temporal averaged amplitude of the tip angle fluctuations, $\langle\hat{\Theta}\rangle$, is shown in Fig. \ref{Fig3MAng} (b). Again, both arrays of flexible filaments show a very similar behaviour on both sides of the D-shaped body. The evolution of $\langle\hat{\Theta}\rangle$ depict a classical Stuart-Landau bifurcation behaviour with a critical reduced velocity close to $U^{*} \simeq 1$, matching the start of the passive reconfiguration observed for the average tip angle, $\langle\Theta\rangle$ in Fig. \ref{Fig3MAng}(a). Below this value, the filament oscillations are very low or practically non-existent ($\langle\hat{\Theta}\rangle<0.2^\circ$), confirming that the flow loading is unable to excite the flexible filaments at those velocities. Then, a sudden increase in the oscillations takes place reaching up to $\langle\hat{\Theta}\rangle \simeq 4^\circ$ for $U^*=1.77$ or $Re=3636$, matching again the saturation phase of $\langle\Theta\rangle$ growth. The attained value can be considered as a saturation of the non-linear coupling between the wake flow and the flexible filaments, since further increasing of the flow velocity ($U^{*}>1.77$) leads to a quasi-saturation zone of the filaments oscillations amplitude. We will analyze further this saturation phase in terms of temporal/frequency locking. However, it is worth mentioning that unlike the average angle,  $\langle\Theta\rangle$, in which the saturated zone had a constant or even decreasing mean deflection, in the case of the filaments oscillations, there is a small increase in the vibration magnitude. For the maximum $U^*$ and $Re$, the flexible filaments deflect with an averaged deformation around $8.6^{\circ}$ along with a vibration of $4.2^\circ$, exhibiting a nearly symmetrical behavior on both sides of the body. Therefore, in the following, we will use the averaged data from both arrays of filaments. 

\subsection{Fluid-structure interaction}\label{subsec:FSI}
\begin{figure}[t]
\centering
\includegraphics[width=1\textwidth]{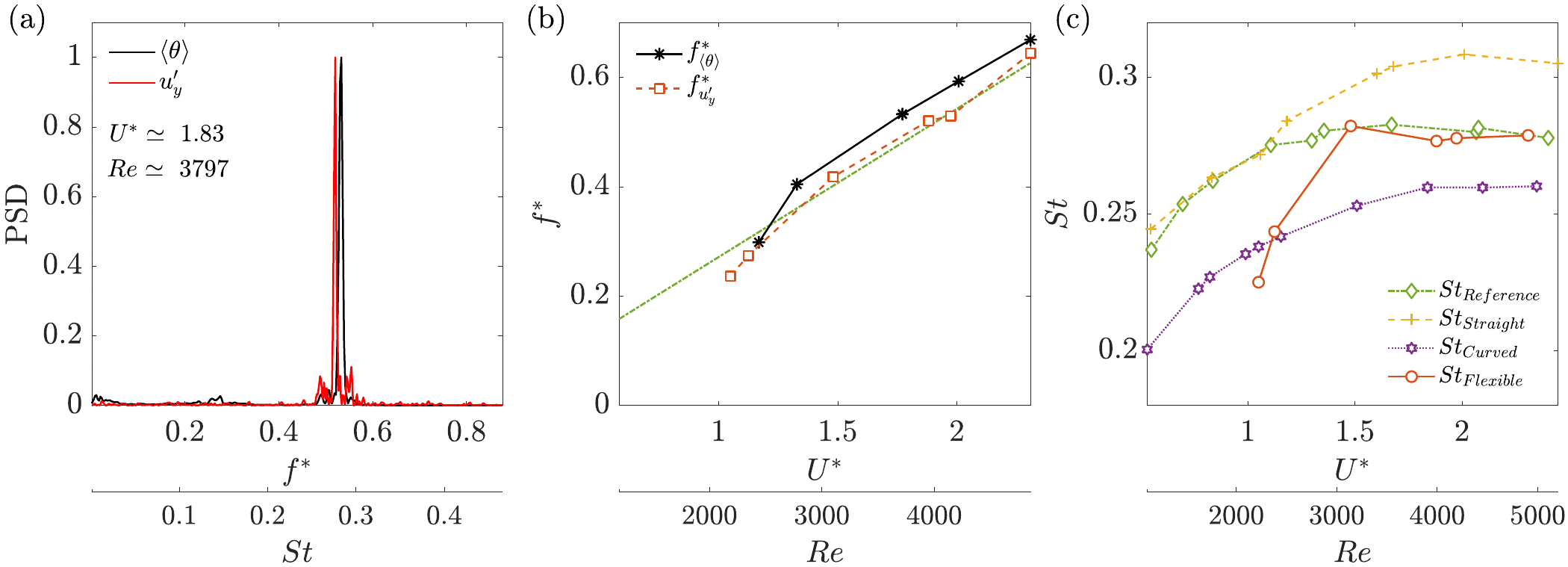}
\caption{(a) PSD spectra of spatial averaged tip angle deflection signal and transverse velocity fluctuations located at (4.25D, -0.4D, 0) for $U^{*}\simeq 1.83$ $(Re\simeq 3797)$. (b) Dominant frequencies of the filament tips' fluctuations ($f^{*}_{\langle\theta\rangle}$) in black solid asterisks. The vortex shedding dominant frequency ($f^{*}_{\langle u_y\rangle}$) for the flexible filaments is included in orange squares. A Strouhal law for the reference case ($St\simeq0.27$, with $f^*=St\cdot U^*$) is also included with a green dashdotted line. (c) Strouhal, $St$, number evolution with $U^{*}$ ($Re$) for vortex shedding in the reference case (green solid line and diamonds); straight rigid filaments (yellow dashed line and crosses); curved rigid filaments (purple dotted line and stars) and flexible filaments (orange solid line and circles) attached to the D-shaped body.}
\label{Fig4FSI}
\end{figure}

In this section, we discuss the connection between the flexible filaments and the vortex shedding behind the D-shaped body. 
The Power Spectral Density (PSD) related to the spatially averaged flexible filament motion, $\langle\Theta\rangle$, and the transverse flow velocity fluctuations, $u'_y$, are presented in Fig. \ref{Fig4FSI} (a). These measurements were taken at the location of maximum transverse velocity fluctuations, specifically at $(4.25D, -0.4D, 0)$, for the flexible filament case, corresponding to $U^* \simeq 1.80$ ($Re\simeq  3800$). The wake information is derived from the temporal evolution signals captured through PIV measurements. At the selected flow conditions, the filaments are passively deflected and vibrating which causes the appearance of an energetic peak around $f^{*}=f/f_n\simeq 0.52$ ($St\simeq 0.28$), indicating a nearly periodic behaviour for the flexible filaments oscillation. In terms of vortex shedding, the velocity probe depicts a frequency peak close to the vibration frequency, which may indicate some synchronization between the flexible filaments and the shed vortices. 

To further analyze this temporal locking, Fig. \ref{Fig4FSI}(b) depicts the trend of the dominant frequencies obtained from the filaments motion, $\langle \Theta \rangle$,  at the tested reduced velocities, $U^{*}$ ($Re$) with black asterisks. Additionally, the vortex shedding dominant frequency is also represented with orange squares. This frequency is obtained from the fluctuating transversal velocity temporal evolution at (4.25D, -0.4D, 0).  Finally, a Strouhal law, corresponding to the averaged dominant vortex shedding frequency of the D-shaped body ($St=0.27$) is introduced in a green dashdotted line in Fig. \ref{Fig4FSI}(b). The associated values of $Re$ (or $St$) are not exactly the same between PIV or visualization measurements as they were recorded separately in slightly different experimental conditions. 
The dominant frequencies of vortex shedding, $f^*_{u'_y}$, and filaments motion, $f^*_{\langle \theta\rangle}$, increase  linearly with $U^{*}$. Both frequencies seem to be locked for $U^{*}>1.13$, the synchronization frequency is a bit higher than the vortex shedding frequency associated to the reference case. The initial synchronization frequency, $f^{*} \simeq 0.3$ coincides with the expected natural frequency of the filaments in water, $f_{n,w} \leq f_{n}/3$, however, as we do not have a exact value of $f_{n,w}$ we have used $f_{n}$ in our analysis. 

A complementary analysis is devoted to the evolution of vortex shedding frequency in terms of Strouhal numbers, that we plot in Fig.\ref{Fig4FSI}(c) for the four tested configurations.
We observe an initial increase of the vortex shedding Strouhal number and then, a saturation phase, which is common given the transitional $Re$ tested in our experiments \citep{Achenbach1981}.

\begin{figure}[t]
\centering
\includegraphics[width=0.8\textwidth]{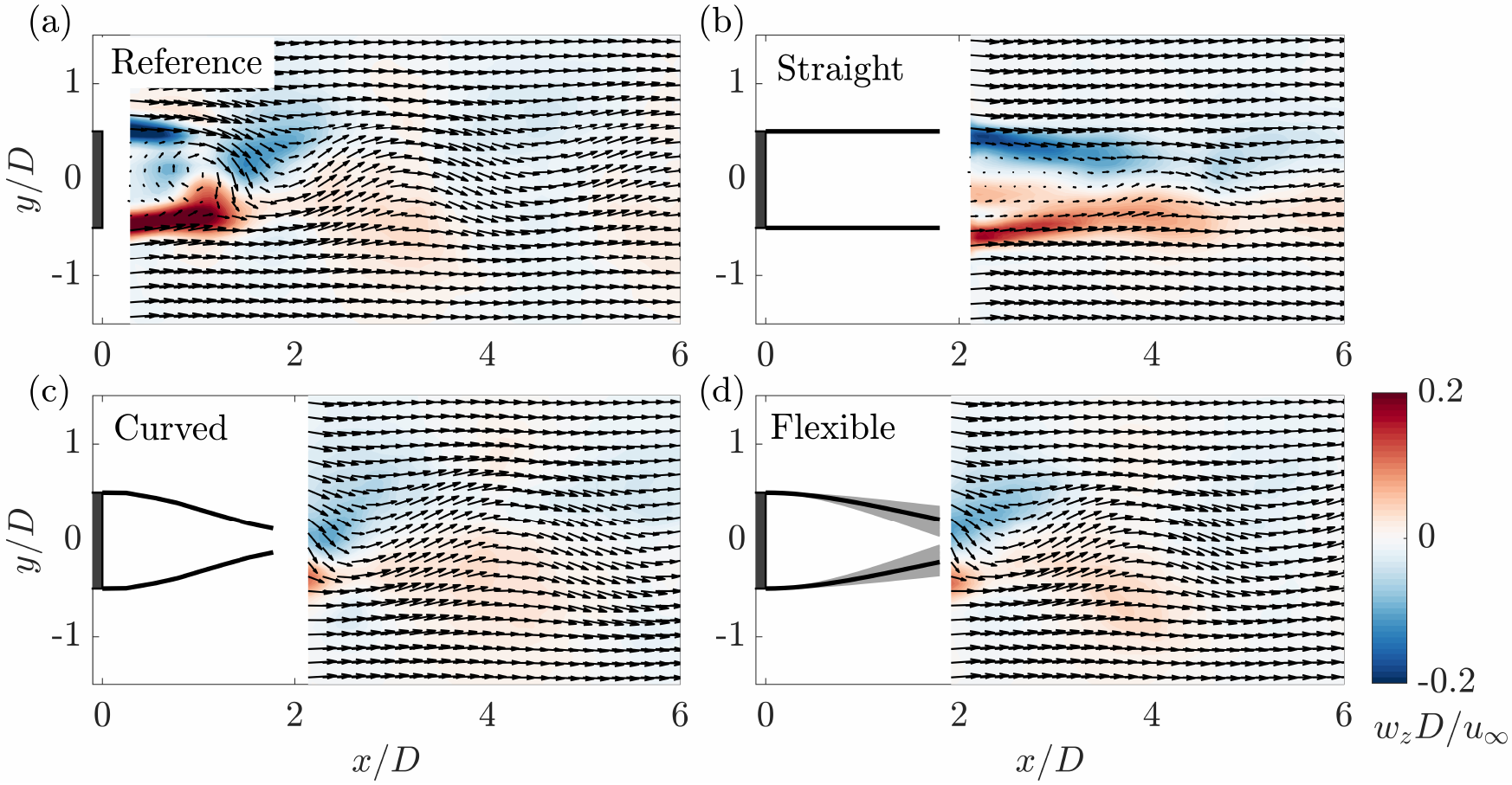}
\caption{Contours of phase-averaged spanwise vorticity ($w_{z}D/u_{\infty}$) and the corresponding flow vectors ($U_x$, $U_y$) at $Re\simeq5015$ ($U^{*}\simeq 2.38$). The four configurations are depicted: (a) Reference case; (b) Straight rigid filaments; (c) Curved rigid filaments and (d) Flexible filaments. In (d) the filament vibrations are represented by a gray area, while the spatio-temporal averaged filament deflection is shown in black. One over two vectors are represented for the flow field.}
\label{Fig5Vorticity}
\end{figure}

When flexible and curved filaments are employed, the vortex shedding frequency is reduced in comparison with the reference case, due to their stabilization effect. In fact, the wake is almost steady for $U^*<1~ (Re < 2052)$ when the flexible filaments are set at the D-shaped body trailing edges. The reduction of $St$ with the filaments might be related to the modification of the flow separation angle. 
The observed change of the shedding frequency for the tested configurations could also indicate different features of the shed vortices. Figure \ref{Fig5Vorticity} illustrates phase-averaged snapshots of spanwise vorticity contours and velocity field flow vectors, for an equivalent shedding phase, for the different configurations. In particular, straight filaments seem to produce more elongated shear layers which will produce smaller transversal velocity fluctuations in comparison with the other cases. In addition, the velocity vectors appear to be aligned with the flow direction for the straight cavity configuration. On the other hand, as previously mentioned, the reduction in body bluffness caused by the curved and flexible filaments cases, leads to a modified flow separation angle as shown in Fig.\ref{Fig5Vorticity} (c,d). The modified separation angle will reduce flow fluctuations in the transverse direction, as the reconfigured shape of the filaments alters the direction of vortex shedding. Additionally, in both cases, the intensity of the vortices generated in the wake is significantly reduced, although the wake is disturbed over a larger region. More importantly, these two configurations are able to remove the recirculation zone in the near wake behind the blunt body.

\subsection{Effects of the different filament configurations on the near wake}\label{subsec:Wakechange}

\begin{figure}[h!]
\centering
\includegraphics[width=1\textwidth]{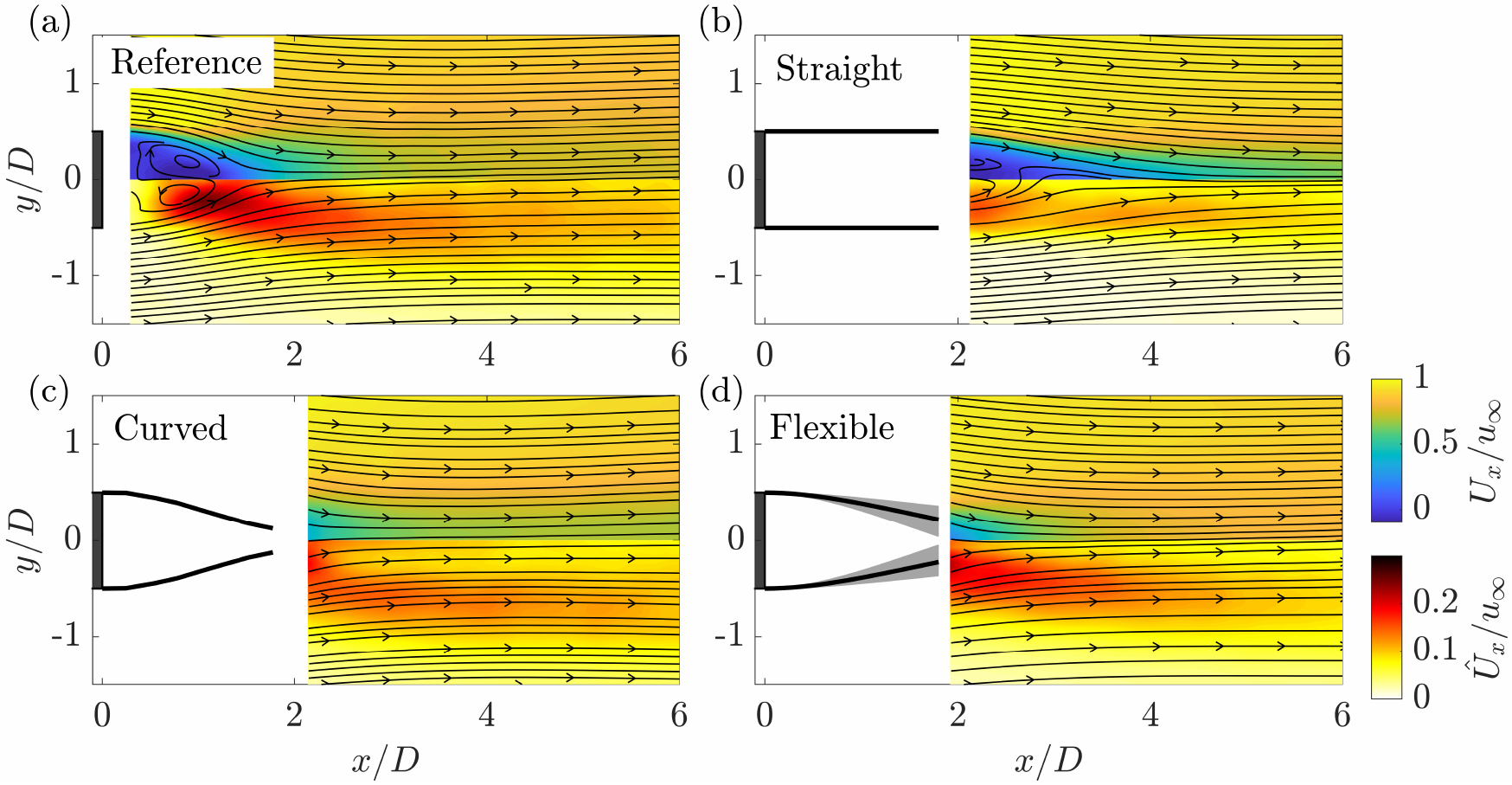}
\caption{Contours of time-averaged of streamwise velocity, $U_x$, (top half of each panel) and contours of the time-averaged streamwise velocity fluctuations, $\hat{U}_x$, (bottom half of each panel) corresponding to: (a) Reference case; (b) Straight rigid filaments; (c) Curved rigid filaments and (d) Flexible filaments configurations. The corresponding averaged flow streamlines ($U_x,U_y$) are also included included. In (d) the filament vibrations are represented by a gray area, while the spatio-temporal averaged filament deflection is shown in black. Data corresponding to $Re\simeq5015$ ($U^{*}\simeq 2.38$).}
\label{Fig6PIVUx}
\end{figure}

Figure \ref{Fig6PIVUx} depicts the time-averaged contours of the streamwise velocity, $U_x$, and the streamwise velocity fluctuations amplitude, $\hat{U}_x$, for the four tested configurations: Reference (a), Straight (b), Curved (c) and Flexible (d) cases. In addition, we depict the corresponding averaged flow streamlines ($U_{x},U_{y}$). At $Re\simeq 5015$ ($U\simeq 2.38$), the flow around the reference D-shaped body produces a near wake composed by two symmetric recirculating zones, whose extension is up to $x\simeq 1.25$. The massive flow separation at the body trailing edges generates vortex shedding provoking important flow fluctuations in the near wake (see Fig.\ref{Fig6PIVUx} a). The extension of that recirculation region is greatly reduced when rigid straight flaps are set, reducing the recirculation zone in size and intensity. Also, the presence of the flaps prevents the flow to enter close to the body base, creating a dead-flow region near the body base, that will increase the pressure there.  The straight flaps are also able to reduce the unsteadiness in the near wake by reducing velocity fluctuations, as shown in Fig. \ref{Fig6PIVUx}(b). When the curved and flexible filaments are mounted, the wake features are pretty similar (see Fig. \ref{Fig6PIVUx}c,d). Both configurations are able to suppress the recirculation regions downstream of the flaps trailing edges. However, they both increase the flow fluctuations in comparison with the Straight case. These devices can alter the separation angle, generating additional flow fluctuations, as noted in \cite{Mariotti2017, Mariotti2019}. While these fluctuations may hinder drag reduction, the elimination of the recirculation region behind the body will ultimately lead to a significant reduction in drag.
Additionally, flexible filaments introduce more unsteadiness in the wake than curved ones, particularly at velocities around $U\sim 2$. This increased excitation of the vortex shedding process may be linked to filament vibrations, illustrated by the gray areas in Fig. \ref{Fig6PIVUx}(d)

\begin{figure}[t]
\centering
\includegraphics[width=1\textwidth]{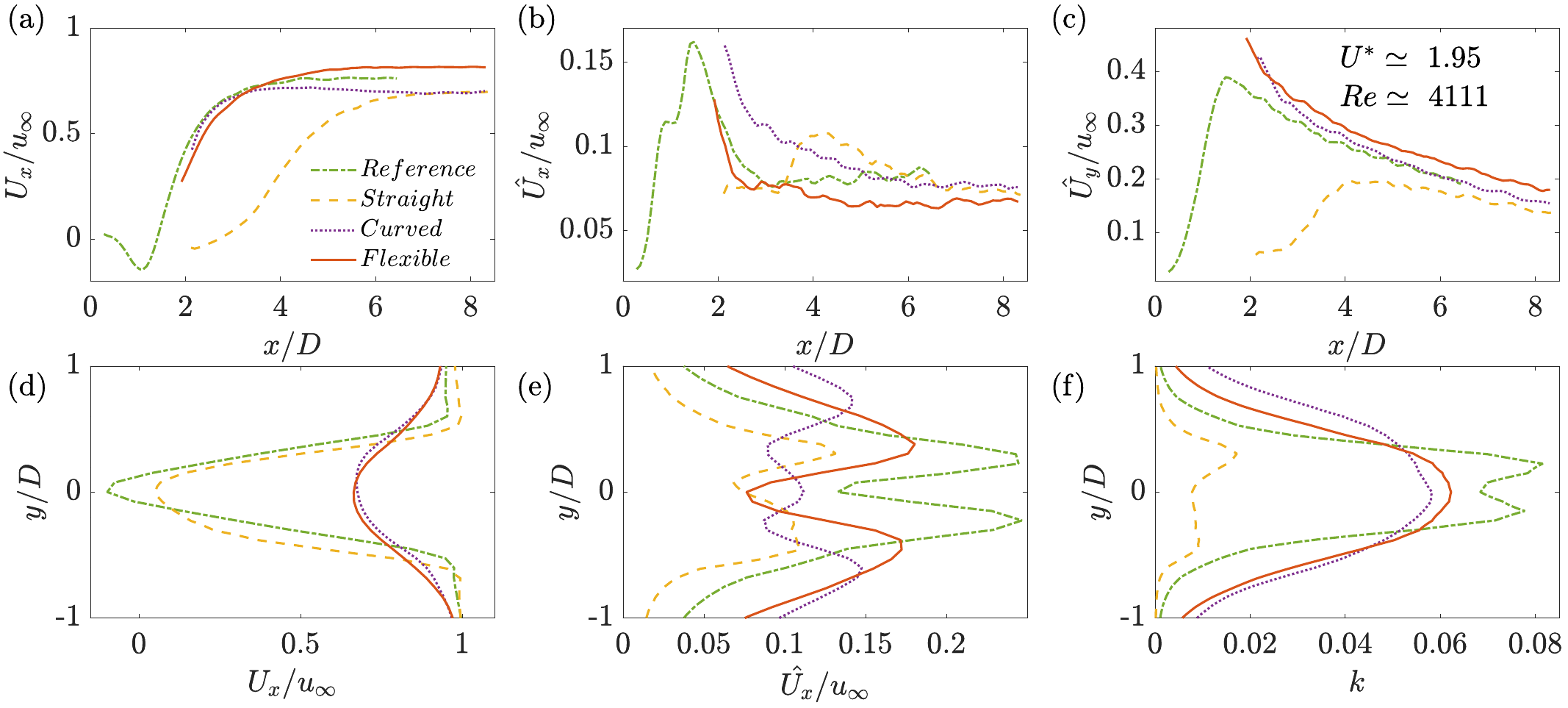}
\caption{(a) Time-averaged streamwise velocity longitudinal profiles, $U_x$. (b) Time-averaged streamwise velocity fluctuations longitudinal profiles, $\hat{U}_x$. (c) Time-averaged transversal velocity fluctuations longitudinal profiles, $\hat{U}_y$. (d) Time-averaged streamwise velocity transversal profiles, $U_x$, (e) Time-averaged streamwise velocity fluctuations transversal profiles, $\hat{U}_x$ and (f) Turbulent kinetic energy transversal profiles $k$. The longitudinal profiles are selected at $y/D=0$. The transversal profiles are placed $x/D\simeq1.26$ for the reference case and at $x/D\simeq3.06$ for the other configurations, at $Re\simeq4111$ ($U^*\simeq1.95$). The four configurations represented are: the Reference case (green solid line); Straight rigid filaments (yellow dashed line); Curved rigid filaments (purple dotted line) and Flexible filaments (orange solid line).}
\label{Fig7GenProfile}
\end{figure}

Furthermore, we analyze the velocity distributions downstream the D-shaped body for the different configurations, focusing on the velocity deficit --- the reduction in $U_x$ due to the influence of the body relative to the $u_{\infty}$, i.e., $U_x/u_{\infty}<1$ --- since it is connected through the momentum balance equation to the hydrodynamic forces acting on the blunt body.
The reduction of recirculation zones and the decrease in the velocity deficit in the wake could indicate a lower drag coefficient. In order to perform that analysis, averaged velocity profiles are represented in Fig. \ref{Fig7GenProfile}. In particular, the axial and cross-sectional velocity profiles for streamwise and transversal velocities (averaged and fluctuations amplitude) are compared for the 4 tested configurations. Furthermore, the turbulent kinetic energy, computed as $k=1/2((\hat{U_{x}}/u_{\infty})^2+(\hat{U_{y}}/u_{\infty})^2)$, transversal profile is also included. The streamwise distribution of $U_x$ at $y=0$ shows quantitatively the effect of the different rear geometries previously depicted in Fig. \ref{Fig6PIVUx}. In comparison with the reference case, the straight filaments attenuate the recirculation region and the velocity gradient outside the recirculation region at $Re\simeq4111$ (see Fig. \ref{Fig7GenProfile}a). Both curved and flexible filaments are able to suppress the negative velocity values, decreasing the velocity deficit downstream the filaments trailing edges. The similarities between these configurations are based on the similar averaged orientation of the flexible filaments ($\langle\Theta\rangle\simeq 9^{\circ}$) in comparison with the curved filaments ($11^{\circ}$) at  $U^*\simeq2.31$. The performed comparison can also be made in terms of streamwise velocity fluctuations, which are mainly associated to vortex shedding in two-dimensional wakes. These flow fluctuations and vortex shedding are related to the generation of unsteady lift force fluctuations and the consequent increase of drag. In that regard, the longitudinal profile of $\hat{U}_{x}$ is shown in Fig. \ref{Fig7GenProfile}(b). The reference case depicts the classical evolution of streamwise velocity fluctuations at $y=0$ for these kind of wakes, with a peak that places the vortex formation length (at $x/D\simeq 1.49$) of the shed vortices \cite{Bearman67}. After that position, the velocity fluctuations decay due to the viscous dissipation of the generated vortices. For the straight configuration, the filaments move away the formed vortices ($x/D\simeq 4.18$) and the associated flow fluctuations are reduced, as it can be seen in Figs. \ref{Fig5Vorticity}, \ref{Fig6PIVUx}. The curved and flexible filaments depict similar trends, illustrating the decay of flow fluctuations observed for the reference case. Interestingly, the extent of the near wake velocity fluctuations in the symmetry plane is considerably reduced with the flexible filaments, reaching a constant value of velocity fluctuations around $x/D\simeq 2$. On the other hand, Fig. \ref{Fig7GenProfile}(c) shows the transverse velocity fluctuations along the flow direction. As previously mentioned in reference to Fig. \ref{Fig6PIVUx}, in both the curved and flexible configurations, the shape of the filaments, adapted to the flow, causes a change in the flow separation direction. This effect increases the level of transverse velocity disturbances since the shedding is no longer occurring parallel to the axial flow direction. As a result, both curves are above the reference case near the body, although the effect diminishes further downstream. However, with the addition of straight rigid filaments, transverse velocity fluctuations are significantly reduced, as mentioned in Sect. \ref{subsec:FSI}.
Regarding the transversal direction, we depict in Fig. \ref{Fig7GenProfile}(d) the streamwise velocity profiles for the tested configurations. The selected streamwise position is $x/D\simeq1.26$  for the reference case (at the maximum flow fluctuations location) and at $x/D =3.06$ for the filament cases (Straight, Curved, Flexible) to maintain the distance between the selected positions and the different trailing edges. The transversal profiles, illustrated in Fig. \ref{Fig7GenProfile}(d), obtained at  $Re\simeq4111$ depict a strong velocity gradient at $y/D \pm 0.5$ for the reference and straight cases. These gradients are reduced when curved and flexible filaments are set, also, the velocity deficit reached by these configurations is also smaller, indicating a modified flow separation angle and an eventually lower drag coefficient. 

Streamwise velocity and turbulent kinetic energy profiles at the mentioned streamwise positions are shown in Figs. \ref{Fig7GenProfile}(e,f). In all cases, the streamwise fluctuations show a double peak in the transverse direction, as evidenced by Figs. \ref{Fig5Vorticity} and \ref{Fig6PIVUx}, indicating the shedding of alternativing vortices detached from each side of the body. The reference case displays the typical transverse distribution of flow fluctuations in the transverse velocity for two-dimensional wakes behind blunt bodies \cite{Wesfreid96}
, with two symmetrical peaks of streamwise velocity fluctuations at the end of the recirculation region (see Fig. \ref{Fig7GenProfile}e). These flow oscillations, in addition to the transverse flow fluctuations, generate a zone of high turbulent kinetic energy, $k$, in the center of the wake (see Fig. \ref{Fig7GenProfile}f) and are responsible for generation of large hydrodynamic forces. Notably, the straight case significantly reduces these flow fluctuations due to the delay of flow separation until the extended trailing edges, generating long shear layers that need to be developed to induce any unsteadiness in the flow (see Fig. \ref{Fig5Vorticity}). Furthermore, the curved and flexible filaments modify the flow separation angle at $x/D=0$, which in turn alters the flow fluctuations downstream. They exhibit similar transverse distributions of flow fluctuations, with intermediate unsteadiness levels between the reference and straight cases. However, the transversal extension of these flow fluctuations is larger than in the other two configurations.
Finally, the flexible filaments generate the largest streamwise velocity fluctuations at $y/D \pm 0.4$, closer to the symmetry axis compared to the other configurations and nearer to the reference case. In this setup, the flow fluctuations are likely driven by the vibrations of the filaments.

\begin{figure}[h!]
\centering
\includegraphics[width=\textwidth]{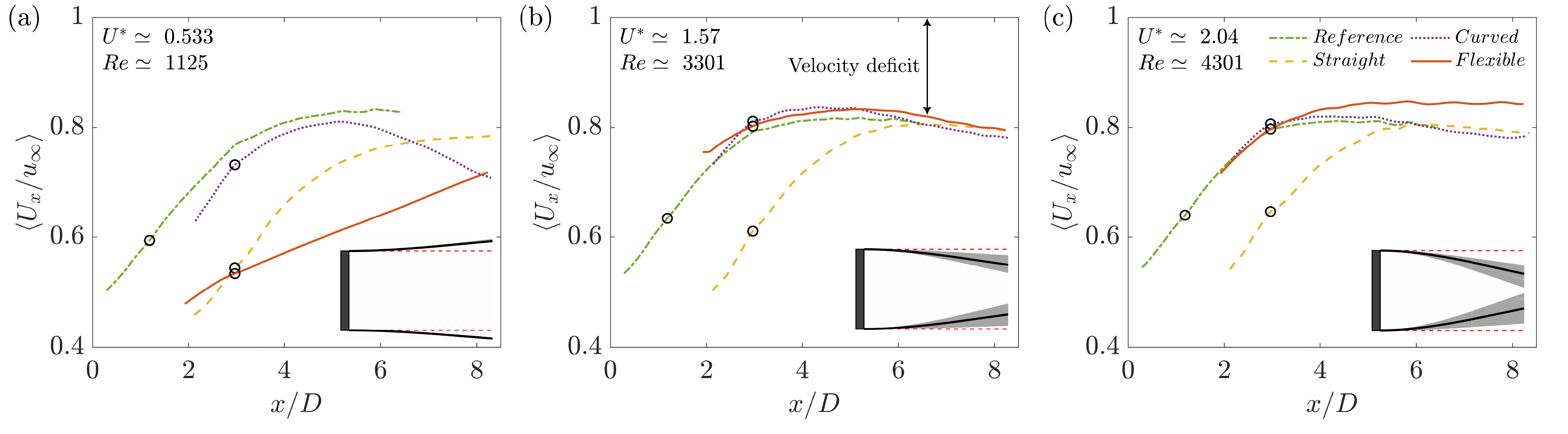}
\caption{Spatio-temporal averaged streamwise velocity evolution, $\langle U_{x} \rangle$, at (a) $Re\simeq1125$ ($U^*\simeq0.533$), (b) $Re\simeq3301$ ($U^*\simeq1.57$) and (c) $Re\simeq4301$ ($U^*\simeq2.04$) for the four tested configurations. The Reference case is represented in green solid dashed-dotted lines; Straight rigid filaments is illustrated by yellow dashed lines; Curved rigid filaments in purple dotted lines and Flexible filaments are depicted in orange solid lines. In each case and configuration, the location where the transversal profiles are depicted is indicated by a black circle (see Fig. \ref{Fig7GenProfile}d-f). For each case, the spatio-temporal averaged deflection of the flexible filaments is shown in an inset by black solid lines, the filament vibrations are depicted with a gray shading, and the reference position of the filaments are indicated by red dashed lines.}
\label{Fig8ProfileRe}
\end{figure}

In order to perform a deeper characterization of the wake, we have computed the spatio-temporal averaged streamwise velocity distribution along the wake to compare the different configurations in terms of velocity deficit and therefore, elucidate the eventual relative drag differences. These velocity distributions are averaged in the range $y/D \pm 1$ for each streamwise location. Therefore, Fig. \ref{Fig8ProfileRe} represents the global streamwise velocity deficit in the near wake of the four tested configurations for three flow conditions, which could correspond to different responses of the flexible filaments and their equivalents in the other configurations.
In Fig.\ref{Fig8ProfileRe}, we can see that the recirculation zone is similar between the reference and straight cases, with nearly parallel velocity deficit distributions between both configurations for the three selected flow conditions, the only difference being an offset corresponding to the $x/D=1.8$ extension of the straight filaments. The curved configuration produces a much smaller velocity deficit than the straight case for all the displayed conditions. Interestingly, there is a velocity reduction further from the base that is not seen for the other configurations. The increasing deformation of the flexible filaments, as $U^{*}$ grows, has a clear effect on the velocity deficit in the wake behind the body. At low flow velocities, $U^{*}\simeq 0.533$, the flow is not able to affect the flexible filaments (see inset in Fig. \ref{Fig8ProfileRe}a). As a result, the pre-existing outward deflection alters the velocity in the wake, displaying a trend that differs from other configurations. When the filament reconfiguration starts, as for example at $U^{*}\simeq 1.57$, the streamwise velocity distribution starts to be similar to other configurations, especially to the curved case. However, the velocity deficit is stronger for the flexible filaments, since the averaged reconfiguration angle is still moderate,  $\langle{\Theta}\rangle \simeq 6^\circ$ (see Fig. \ref{Fig8ProfileRe}b). In the saturated phase of response, $U^{*}>1.77$, both flexible and curved configurations shows a very similar velocity deficit in the near wake. However, the velocity decrease observed in the far wake for the curved case is not seen in the flexible case, which may indicate an advantageous reduction of the drag coefficient due to passive, dynamic reconfiguration. 

\begin{figure}[h!]
\centering
\includegraphics[width=0.8\textwidth]{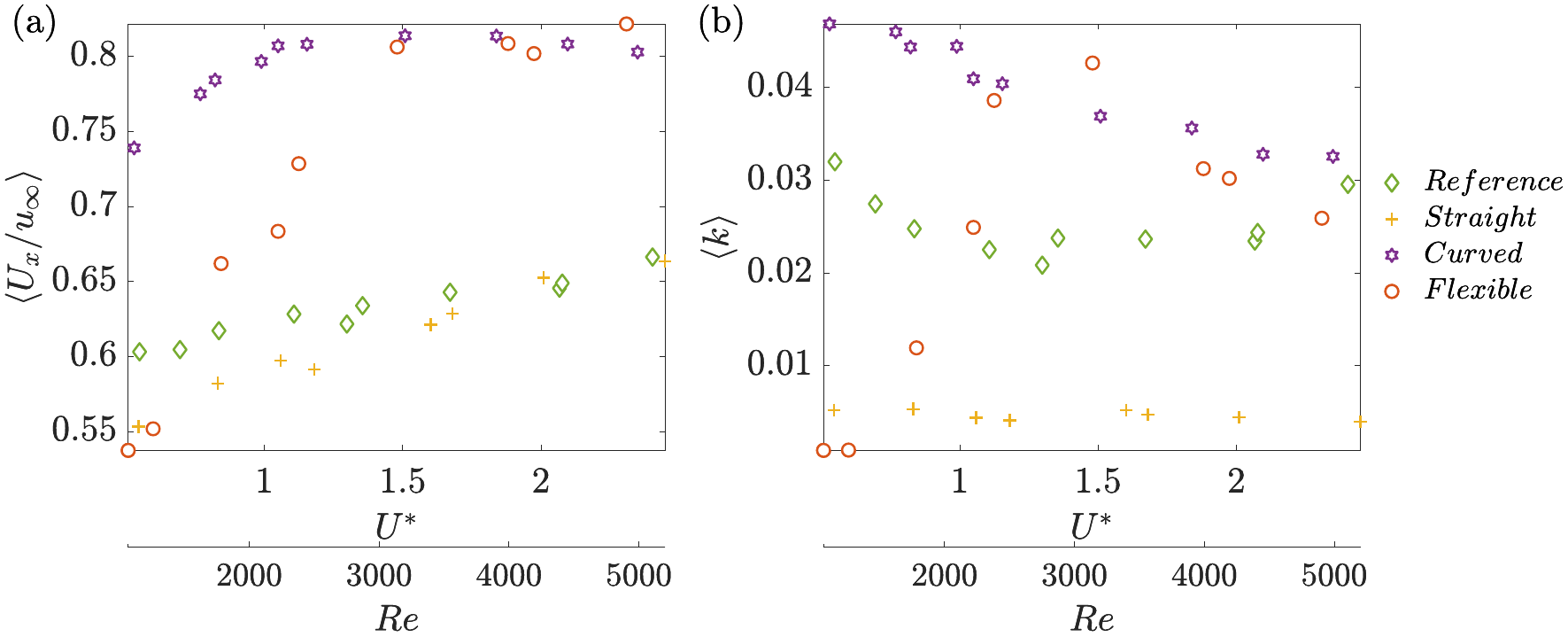}
\caption{Evolution of the (a) spatio-temporal averaged streamwise velocity, $\langle{U_{x}}\rangle$, and (b) spatio-temporal averaged turbulent kinetic energy, $\langle k\rangle$, with $Re$ ($U^*$). The four configurations represented are: the Reference case (green diamonds); Straight rigid filaments (yellow crosses); Curved rigid filaments (purple stars) and Flexible filaments (orange circles). Note that the spatial averaging has been performed over $y/D \pm 1$  at $x/D\simeq 1.26$ for the reference case and at $y/D \pm 1$ and $x/D\simeq3.06$ for the other configurations.} 
\label{Fig9MFlucUx}
\end{figure}

To further analyze the wake modifications induced by the different devices, we selected a streamwise position of $x/D\simeq3.06$ ($x/D\simeq1.26$ for the reference case) to compare the spatio-temporal averaged streamwise velocity and turbulent kinetic energy. Figure \ref{Fig9MFlucUx} compares the impact of various flow conditions on the performance of each rear device using the spatio-temporal averaged streamwise velocity distribution (both averaged and fluctuating components) over $y/D \pm 1$. By maintaining the same relative position to the trailing edges of the filaments, we can directly compare the tested devices. The reference case depicts a slight decrease of the velocity deficit for the studied range of $Re$ numbers (see Fig. \ref{Fig9MFlucUx}a), indicating the growth of the near wake. The straight filaments show a similar trend, with only small differences at low flow velocities. The curved filaments, however, are able to reduce importantly the velocity deficit and also depict the decreasing trend of the velocity deficit with $Re$.  Regarding the flexible filaments, the evolution of the velocity deficit illustrates the response of the filaments depending on the flow conditions as shown in Sect. \ref{subsec:Response}. At low reduced velocities, $U^{*}<1$, the velocity deficit is quite similar to the one corresponding to the rigid filaments, as the flow is not able to move or deform the filaments. Then, the velocity deficit is decreased as a sign of the passive reconfiguration experienced between $1<U^{*}<1.77$. Finally, in the saturation regime, $U^{*}>1.77$, the curved and flexible flaps have very similar velocity deficit and much smaller than the one corresponding to the reference or straight cases. 

Focusing now on the flow fluctuations, estimated here with the spatio-temporal average of the turbulent kinetic energy over $y/D \pm 1$, $\langle{k}\rangle$,  Fig. \ref{Fig9MFlucUx} (b) shows the evolution for the four tested configurations. For the reference case, the flow unsteadiness seems to converge around $Re \simeq 1755$, as it happened with the shedding frequency (see Fig. \ref{Fig4FSI}c). As previously observed, straight filaments significantly reduce wake unsteadiness across all tested $Re$, maintaining a constant fluctuation amplitude throughout the entire range. Moreover, curved filaments provoke a wider near wake (as seen in Fig. \ref{Fig8ProfileRe}c) which produces larger spatio-temporal flow fluctuations in comparison with other configurations. The distinct behavior of the flexible filaments is also evident in Fig. \ref{Fig9MFlucUx}(b). At low reduced velocities, flow fluctuations are minimal, as the flow is insufficient to displace the filaments or generate a significant near wake. As the flow velocity increases, fluctuations sharply rise, coinciding with the increasing passive reconfiguration and vibration of the filaments. However, at highest reduced velocities, $U^{*}> 1.77$, the flexible filaments behave similarly to the curved configuration, with minimal impact from filament vibrations on wake unsteadiness. 

\section{Conclusions}\label{sec:Conclusions}
In this study, we experimentally compared the effect of 4 different afterbody configurations in the flow around a two-dimensional D-shaped blunt body.  The main goal of the work was to asses the efficiency of rear flexible filaments to reduce the wake extent and the flow unsteadiness behind the body, which may contribute to the generation of undesirable forces in several applications. This problem is a simplified configuration of many associated industrial and environmental problems. The experiments, performed in a closed-loop water channel, were conducted at different flow velocities to analyze the relation between the tested afterbodies and the corresponding near wake. In the case of the flexible filaments, they are seen to interact with the flow around the body in different regimes (see Fig. \ref{Fig3MAng}). At small flow velocities, $U^{*}<1$, the filaments are not affected by the flow and they preserve their initial loading. When the flow velocity is increased $1<U^{*}<1.77$, the flexible filaments start to passively reconfigure as well as displaying mild vibrations (see Fig. \ref{Fig3MAng}). The response of the flaps saturates at $U^{*}>1.77$, fixing an averaged tip deflection deformation of $\langle\Theta\rangle \simeq 9^\circ$ associated to $\langle\hat{\Theta}\rangle \simeq 4^\circ$ vibrations. This saturated passive reconfiguration is selected to produce the curved filament configuration. The response of the filaments is almost two-dimensional (see Fig. \ref{Fig2AngZ}) and it is produced by a temporal locking with the vortex shedding produced in the wake behind the model (see Fig. \ref{Fig4FSI}). This synchronization provokes the reduction of the shedding frequency when this flexible filaments are implemented. Regarding the wake changes induced by the different afterbodies, in comparison with the reference case, flexible and rigid (straight or curved) filaments delay the flow separation, increasing the distance between the recirculating flow or velocity fluctuations and the body base, eventually increasing the base pressure and reducing the unsteady loads on the body (see Fig. \ref{Fig5Vorticity}, \ref{Fig6PIVUx}, \ref{Fig7GenProfile}). More quantitatively, the streamwise velocity deficit in the wake, an indirect indicator of the associated drag, is reduced when the filaments are passively reconfigured (see Fig. \ref{Fig8ProfileRe}b,c). Therefore, when the flexible filaments interact with the flow $U^{*}>1$, the effect of them in the wake passes from the straight filaments to the curved ones (see Fig. \ref{Fig8ProfileRe}, \ref{Fig9MFlucUx}), due to their mean deformation. In addition, their vibrating amplitude generates a small influence on the wake features, as the vibrating flexible filaments show very similar results compared to the rigid curved ones. These results prove the potential of flexible filaments as efficient solution for aerodynamic improvement through passive reconfiguration, without the disadvantages that may feature continuous flexible parts (such as significant loads and potential torsional deformations). That said, its application to engineering problems may require a deeper analysis under more complex geometries and flow conditions.
\begin{acknowledgments}

The authors would like to thank Alejandro Ibarra, Jean-Fran\c cois Egea, Amaury Fourgeaud, Xavier Benoit-Gonin, Olivier Brouard, Laurent Quartier and Jose Eduardo Wesfreid for their support in the preparation of the experiments and the establishment of this international scientific collaboration. The authors gratefully acknowledge the funding provided by the projects TED2021-131805B-C21, TED2021-131805B-C22 and PID2022-140433NA-I00 financed by the Spanish MCIN/ AEI/10.13039/501100011033/,  the European Union NextGenerationEU/PRTR and FEDER, UE respectively. J.C.M.H. also acknowledges for the support of the Spanish MECD through FPU20/07261 and EST23/00743.
\end{acknowledgments}

\section*{References}
\bibliography{FlexibleFilaments}

\end{document}